\documentclass[conference]{IEEEtran}
\let\labelindent\relax
\usepackage{amsmath,amssymb,amsfonts,chemarrow,balance}
\usepackage{hyperref}
\usepackage{amsmath} 
\usepackage{graphicx}
\usepackage{subfigure}
\usepackage{url}
\usepackage{mathenv}
\usepackage{color}
\usepackage{breqn}
 \usepackage{algpseudocode}
\newcommand{\BfPara}[1]{{\noindent\bf#1.}\xspace}
\usepackage{multirow}
\usepackage{float}
\usepackage{threeparttable}
\usepackage[inline]{enumitem}

\usepackage[T1]{fontenc}
\usepackage[utf8]{inputenc}
\usepackage{environ}
\usepackage{tikz}
\usetikzlibrary{arrows}
\usetikzlibrary{calc,matrix}
\renewcommand{\baselinestretch}{0.99} 
{\bfseries}{\itshape}
{\bfseries}{\itshape}
{\bfseries}{\itshape}
{\bfseries}{\itshape}
{\bfseries}{\itshape}
{\itshape}

\usepackage{epsfig}

\usepackage{xspace}
\usepackage{setspace}
\newcommand{\note}[1]{}

\newcommand{\etc}{{etc.}\xspace}
\newcommand{\eg}{{\em e.g.}\xspace}

\newcommand{\etal}{{\em et al.}\xspace}

\usepackage{booktabs}

\usepackage{cite}
\usepackage{hyperref}
\definecolor{linkcolour}{rgb}{0,0.2,0.6}
\definecolor{xgreen}{rgb}{0.2,0.6,0.0}
\definecolor{xred}{rgb}{0.7,0.1,0.0}
\hypersetup{  
	pdfpagemode=pagewidth,
	plainpages=false, 
	colorlinks,  
	urlcolor=black,   
	linkcolor=black,    
	citecolor=black,  
	bookmarksnumbered
}
\def\equationautorefname~#1\null{(#1)\null}

\usepackage{listings}
\colorlet{punct}{red!60!black}
\definecolor{background}{HTML}{ffffff }
\definecolor{delim}{RGB}{20,105,176}
\colorlet{numb}{magenta!60!black}

\definecolor{light-gray}{gray}{0.95}
\definecolor{darkgray}{rgb}{0.4, 0.4, 0.4}
\definecolor{editorGray}{rgb}{0.95, 0.95, 0.95}
\definecolor{editorOcher}{rgb}{1, 0.5, 0} % #FF7F00 -> rgb(239, 169, 0)
\definecolor{editorGreen}{rgb}{0, 0.5, 0} % #007C00 -> rgb(0, 124, 0)
\definecolor{orange}{rgb}{1,0.45,0.13}      
\definecolor{olive}{rgb}{0.17,0.59,0.20}
\definecolor{brown}{rgb}{0.69,0.31,0.31}
\definecolor{purple}{rgb}{0.38,0.18,0.81}
\definecolor{lightblue}{rgb}{0.1,0.57,0.7}
\definecolor{lightred}{rgb}{1,0.4,0.5}
\usepackage{upquote}
\usepackage{listings}

\definecolor{pblue}{rgb}{0.13,0.13,1}
\definecolor{pgreen}{rgb}{0,0.5,0}
\definecolor{pred}{rgb}{0.9,0,0}
\definecolor{pgrey}{rgb}{0.46,0.45,0.48}

\makeatletter
\let\matamp=&
\catcode`\&=13
\makeatletter
\def&{\iftikz@is@matrix
  \pgfmatrixnextcell
  \else
  \matamp
  \fi}
\makeatother

\newcounter{lines}

\newcounter{vtml}
\newif\ifvtimelinetitle
\newif\ifvtimebottomline
\tikzset{description/.style={
  column 2/.append style={#1}
 },
 timeline color/.store in=\vtmlcolor,
 timeline color=red!80!black,
 timeline color st/.style={fill=\vtmlcolor,draw=\vtmlcolor},
 use timeline header/.is if=vtimelinetitle,
 use timeline header=false,
 add bottom line/.is if=vtimebottomline,
 add bottom line=false,
 timeline title/.store in=\vtimelinetitle,
 timeline title={},
 line offset/.store in=\lineoffset,
 line offset=4pt,
}

\NewEnviron{vtimeline}[1][]{%
\setcounter{lines}{1}%
\stepcounter{vtml}%
\begin{tikzpicture}[column 1/.style={anchor=east},
 column 2/.style={anchor=west},
 text depth=0pt,text height=1ex,
 row sep=1ex,
 column sep=1em,
 #1
]
\matrix(vtimeline\thevtml)[matrix of nodes]{\BODY};
\pgfmathtruncatemacro\endmtx{\thelines-1}
\path[timeline color st] 
($(vtimeline\thevtml-1-1.north east)!0.5!(vtimeline\thevtml-1-2.north west)$)--
($(vtimeline\thevtml-\endmtx-1.south east)!0.5!(vtimeline\thevtml-\endmtx-2.south west)$);
\foreach \x in {1,...,\endmtx}{
 \node[circle,timeline color st, inner sep=0.15pt, draw=white, thick] 
 (vtimeline\thevtml-c-\x) at 
 ($(vtimeline\thevtml-\x-1.east)!0.5!(vtimeline\thevtml-\x-2.west)$){};
 \draw[timeline color st](vtimeline\thevtml-c-\x.west)--++(-3pt,0);
 }
 \ifvtimelinetitle%
  \draw[timeline color st]([yshift=\lineoffset]vtimeline\thevtml.north west)--
  ([yshift=\lineoffset]vtimeline\thevtml.north east);
  \node[anchor=west,yshift=16pt,font=\large]
   at (vtimeline\thevtml-1-1.north west) 
   {\textsc{Timeline \thevtml}: \textit{\vtimelinetitle}};
 \else%
  \relax%
 \fi%
 \ifvtimebottomline%
   \draw[timeline color st]([yshift=-\lineoffset]vtimeline\thevtml.south west)--
  ([yshift=-\lineoffset]vtimeline\thevtml.south east);
 \else%
   \relax%
 \fi%
\end{tikzpicture}
}

\begin{document}
\title{Analyzing Endpoints in the Internet of Things Malware}

 
\author{\IEEEauthorblockN{Jinchun Choi$^{\diamond\ddagger\star}$, Afsah Anwar$^{\diamond\star}$, Hisham Alasmary$^\diamond$, \\ Jeffrey Spaulding$^\dagger$, DaeHun Nyang$^\ddagger$ and Aziz Mohaisen$^\diamond$} \\
\IEEEauthorblockA{$^\diamond$University of Central Florida \hspace{3mm} $^\dagger$Niagara University \hspace{3mm} $^\ddagger$Inha University\hspace{3mm} $^\star$Equal contributors \\ \{jc.choi, afsahanwar, hisham\}@knights.ucf.edu, jspaulding@niagara.edu \\ nyang@inha.ac.kr, amohaisen@gmail.com}
}

\maketitle

\begin{abstract}
The lack of security measures in the Internet of Things (IoT) devices and their persistent online connectivity give adversaries an opportunity to target them or abuse them as intermediary targets for larger attacks such as distributed denial-of-service (DDoS) campaigns. In this paper, we analyze IoT malware with a focus on endpoints
to understand the affinity between the dropzones and their target IP addresses, and to understand the different patterns among them. Towards this goal, we reverse-engineer 2,423 IoT malware samples to obtain IP addresses. We further augment additional information about the endpoints from Internet-wide scanners, including Shodan and Censys. We then perform a deep data-driven analysis of the dropzones and their target IP addresses and further examine the attack surface of the target device space.

\end{abstract}

\section{Introduction}\label{sec:introduction}
With the number of seamlessly connected and online IoT devices soaring into the 10's of billions~\cite{Middleton2017}, potential adversaries 
set such devices on target via malicious codes. Such codes or malware not only infect the devices themselves but also receive updates from their Command and Control (C2)/dropzones around the world.
Forming a network, these devices have the potential to launch attacks on other targets resulting in distributed denial-of-service (DDoS) attacks~\cite{7971869}.

Reckoning that the malware sources, C2 servers, the intermediary targets, and the victim must be connected to the Internet for the attacks to happen, studying these endpoints is important. This work attempts to understand such endpoints to decipher such patterns. In particular, we extract endpoints from IoT malware samples and perform a data-driven analysis to understand geographical affinities and their exposure to risk. 

Our work is important given its insight into understanding the Indicators of Compromise (IoCs), and the behavioral aspects necessary for threat hunting. Specifically, we make the following contributions: 
\begin{enumerate*}
    \item IP Centric Analysis: We investigate the target IP addresses among different dropzone IP addresses. Additionally, we analyze the locations of dropzones and their target IP addresses. Moreover, we analyze the risk associated with the IP addresses through insights gained from Shodan~\cite{shodan}.
    \item Network Centric Analysis: For the masked target endpoints, we examine the entire network and study the network devices and their exposure to risk.
\end{enumerate*}

\section{Dataset Creation}\label{sec:dataGoal}

Our primary dataset contains a total of 2,423 IoT malware samples collected from IoTPOT~\cite{Pa2016_IoTPOT}. 
We design a tool to reverse-engineer each of the malware samples and extract the IP addresses from the malware code base. 
In total, we extract a total of 106,428 target IP addresses, resulting in 2,211 \textit{unique target} IP addresses associated with 973 malware samples. We also find a total of 2,407 IP addresses, resulting in 877 \textit{unique dropzone} (source of attack)/C2 IP addresses corresponding to 2,318 malware samples.
We notice that only 40\% of the 2,423 malware samples contain target IP addresses, and a total of 95.66\% of 2,423 malware samples contain dropzone IP addresses.
Additionally, we augment the IP addresses with other details including their:  country, city, Autonomous System Number (ASN), and location, using online DNS and IP lookup tools like the \textit{UltraTools}~\cite{ultratools}.
Furthermore, we utilize Shodan~\cite{shodan} to obtain the vulnerable endpoints.

\section{IP centric analysis}\label{sec:ipCentric}

\subsection{Dropzone-Target Inter-relationship}
We examine the affinity between the dropzone and the target IP addresses and find that $\approx$77\% of the unique target IPs receive less than 10 attacks, while
we see that one unique target IP received 72 attacks.
We also find a dropzone IP associated with one malware targeting 1,265 target IP addresses, which is significantly larger than the average (121).

\begin{figure}[t]
	\centering
	\includegraphics[width=1.0\linewidth]{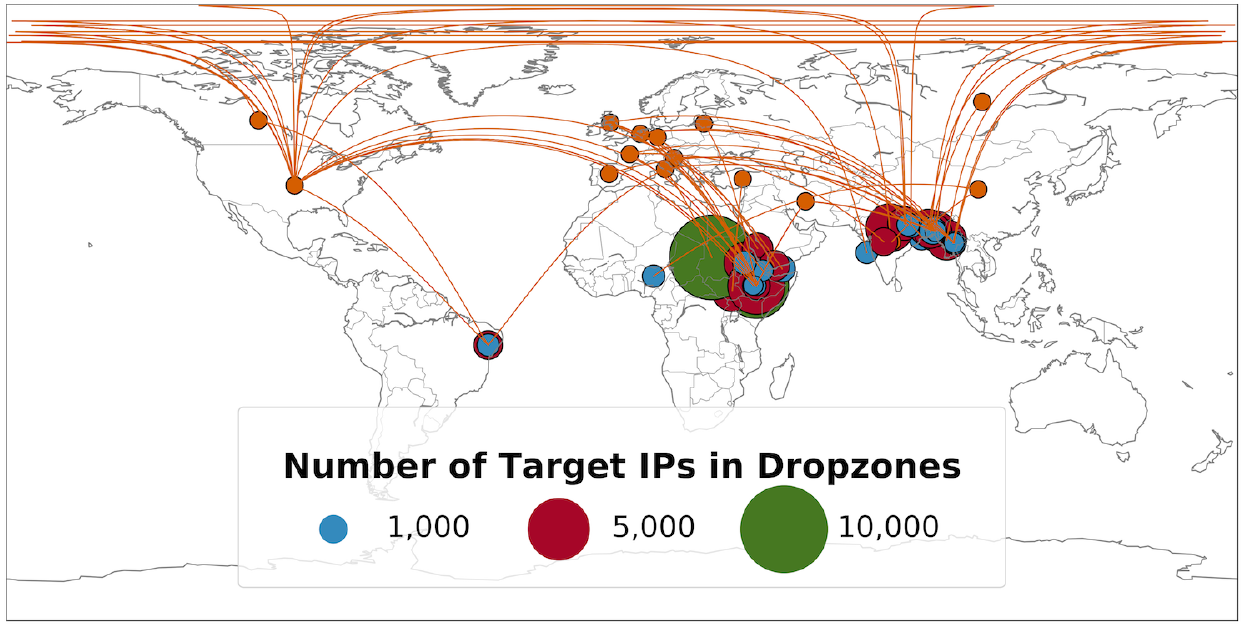}\vspace{-2mm}
	\caption{Attack trends between dropzones and target IPs corresponding to dropzones having over 500 target IPs. The orange circle represents dropzones, and blue, red, and green circles stand for target areas.}\vspace{-2mm}
	\label{fig:trend_attack}
\end{figure}

\BfPara{Shared targets between dropzones}
To inspect the shared targets between dropzone IP addresses, we group the dropzone IP addresses and capture the common targets among the dropzones.
There were 2,199 cases (12.11\%) with 100\% overlap between dropzones. Overall, we found 6,451 cases (35.53\%), with $\textgreater$80\% overlap, and 886 cases (4.88\%) with $\textless$10\% overlap. 
If the target IP addresses between different dropzones are identical, it is possible that the attacker obtained the same targets through similar vulnerability analysis (\eg Shodan) or shared the target list from other attackers through underground communities.

\vspace{-2mm}
\subsection{Geographical analysis}
In this section, we focus on the distribution of the distances between the dropzones and their target IPs. To visualize the flow of attacks in a holistic sense, we plot circular areas whose sizes are proportional to the number of targets placed according to their \textit{average} position (not the exact position) on a world map with geodesic lines originating from various dropzone locations. \autoref{fig:trend_attack} shows the country-level perspective of the relationship.
We observe a large concentration of target areas focused in Central Asia.

The distribution of dropzone and target IPs by country is shown in \autoref{tab:country}. We notice a large distribution of dropzones from US pointing to targets in Asian countries such as Vietnam. Additionally, China and Brazil are target of attacks originating from European countries. 
Imperva Incapsula states that Vietnam (12.8\%), Brazil (11.8\%) and China (8.8\%) were the countries with the most-infected devices (from the Mirai botnet)~\cite{impervaMirai}.

\begin{table}[t]
\small
\caption{Top 5 number of target and dropzone IPs by country. {\upshape Countries include: United States (US), Netherlands (NL), France (FR), United Kingdom (GB), Italy (IT), Vietnam (VN), Brazil (BR), China (CN), India (IN), and Pakistan (PK).}}\vspace{-2mm}
\label{tab:country}
\centering
\begin{tabular}{lccccc}
\toprule
\footnotesize Country & \scriptsize \# Dropzones & \footnotesize \%Total & \footnotesize Country & \scriptsize \# Targets & \footnotesize \%Total \\ \hline
US               & 1,041                & 43.25       & VN               & 26,290             & 24.70       \\ %\hline
NL               & 278                 & 11.55       & BR               & 20,572             & 19.33       \\ %\hline
FR               & 188                 & 7.81        & CN               & 15,799             & 14.84       \\ %\hline
GB               & 183                 & 7.60        & IN               & 5,598              & 5.26        \\ %\hline
IT               & 177                 & 7.35        & PK               & 5,076              & 4.77        \\ %\hline
\bottomrule
\end{tabular}\vspace{-3mm}
\end{table}

\vspace{-2mm}

\subsection{Network Penetration Analysis}

This section focuses on analyzing attributes gathered from Shodan and Censys~\cite{censys}, namely, active ports, vulnerabilities.

\BfPara{Active Ports}
For each dropzone and target IP address, we use information gathered from Shodan and Censys the list of active ports. We extracted 5,745 active ports from 716 of 877 dropzone IPs and 1,114 active ports from 129 of 189 non-masked target IPs. 
Each port number is typically associated with a service, such as port 80 for HTTP traffic.
While we observe common services like SSH (port 22), HTTP (port 80), and HTTPS (port 443), we point out to the usage of Network Time Protocol (NTP) on port 123. NTP is UDP-based and can be prone to ``IP spoofing'' for DDoS attacks~\cite{graham-cumming2014}. This attack is also emphasized in~\cite{verisignUdpFlood}, since the attacker can amplify attack packet size 1,000 larger by exploiting NTP.

\BfPara{Vulnerabilities}
We then examine the susceptibility of the IP addresses; in particular, we determine the vulnerabilities present in the IP addresses. We gather the Common Vulnerabilities and Exposures (CVE) identifier, maintained by MITRE~\cite{mitreCve}.
In our analysis, the second-most common CVE among the dropzones: CVE-2014-1692. The NVD~\cite{nvd} reports the severity of this vulnerability as ``high'' since it might allow remote attackers to cause a Denial of Service (DoS) through memory corruption due to uninitialized data structures from the \texttt{hash\_buffer} function in OpenSSH.

\section{Network centric analysis}\label{sec:networkCentric}
We observe the malware targeting multiple IoT devices for propagation. In this regard, they often mask the endpoints on target. For such endpoints, we analyze their CIDR network. In total, we inspect 27 unique /24, 435 unique /16, and 125 unique /8 IP addresses. Towards this, we evaluate and analyze these networks to investigate their exposure to risk. In particular, we examine the devices on these networks and looked at the services being used.

The corresponding 100,793,403 active IP addresses are then clustered by their device type. Considering that open ports lead to increased security risks, we look for ports that are necessary for a device to operate uninterrupted. Taking a conservative approach, we suggest that if a port is being used by less than 10\% of devices in a given device type, it should be closed to reduce its exposure to risk. Our results show the susceptibility of high-wattage IoT devices, such as heating, ventilation, and air conditioning (HVAC), power distribution units (PDU), \etc, can be abused by the attackers to launch large-scale coordinated attacks. Additionally, the high presence of such susceptible devices lays the foundation for attacks as demonstrated by Soltan \etal~\cite{SoltanMP18}.

\section{Concluding Remarks}\label{sec:conclusion}

In this paper, we analyze the $\approx$78.2\% of total responsive public IPv4 endpoints among dropzones and their targets as extracted from IoT malware and spread across the globe from diverse perspectives. Additionally, we augment our analysis results by leveraging the use of IoT search engines like Shodan or Censys.

\section*{Acknowledgment}
This work was supported by NSF CNS-1809000, NRF-2016K1A1A2912757, and collaborative seed grant from the Florida Cybersecurity Center (FC2).

\balance
\bibliographystyle{IEEEtran}
\bibliography{ref,conf}

\end{document}